\documentclass[twocolappendix,appendixfloats,]{emulateapj}

\usepackage[colorlinks = true, linkcolor = blue, urlcolor  = blue, citecolor = blue, anchorcolor = blue]{hyperref}
\usepackage{hyperref}
\usepackage{apjfonts}
\usepackage{rotating}
\usepackage{amsmath}
\usepackage{color}
\usepackage{color}
\usepackage{CJK}

\newcommand{\thetae}{\theta_{\rm E}}
\newcommand{\pie}{\pi_{\rm E}}

\definecolor{darkbrown}{RGB}{139,69,19}

\shorttitle{MOA-2016-BLG-319L\lowercase{b}}
\shortauthors{Han et al.}

\begin{document}

\title{MOA-2016-BLG-319L\lowercase{b}:
Microlensing Planet Subject to Rare Minor-Image Perturbation Degeneracy in Determining Planet Parameters
}

\author{
Cheongho~Han\altaffilmark{A01}, Ian~A.~Bond\altaffilmark{A02,102}, Andrew~Gould\altaffilmark{A03,A04,A05,101}, \\
and \\
Michael~D.~Albrow\altaffilmark{A06}, Sun-Ju~Chung\altaffilmark{A03,A07}, Youn~Kil ~Jung\altaffilmark{A03}, 
Kyu-Ha~Hwang\altaffilmark{A03}, Chung-Uk~Lee\altaffilmark{A03}, Yoon-Hyun~Ryu\altaffilmark{A03},
In-Gu~Shin\altaffilmark{A08}, Yossi~Shvartzvald\altaffilmark{A09}, Jennifer~C.~Yee\altaffilmark{A08},
Sang-Mok~Cha\altaffilmark{A03,A10}, Dong-Jin~Kim\altaffilmark{A03}, Hyoun-Woo~Kim\altaffilmark{A03}, 
Seung-Lee~Kim\altaffilmark{A03,A07}, Dong-Joo~Lee\altaffilmark{A03}, Yongseok~Lee\altaffilmark{A03,A10}, 
Byeong-Gon~Park\altaffilmark{A03,A07}, Richard~W.~Pogge\altaffilmark{A04}, Chun-Hwey~Kim\altaffilmark{A11} \\ 
(The KMTNet Collaboration),\\
Fumio~Abe\altaffilmark{A12}, Richard Barry\altaffilmark{A13}, David~P.~Bennett\altaffilmark{A13,A14},
Aparna~Bhattacharya\altaffilmark{A13,A14}, Martin~Donachie\altaffilmark{A15}, Akihiko~Fukui\altaffilmark{A16}, 
Yuki~Hirao\altaffilmark{A17}, Yoshitaka~Itow\altaffilmark{A12}, Kohei~Kawasaki\altaffilmark{A17},
Iona~Kondo\altaffilmark{A17}, Naoki~Koshimoto\altaffilmark{A18,A19}, Man~Cheung~Alex~Li\altaffilmark{A15},
Yutaka~Matsubara\altaffilmark{A12}, Yasushi~Muraki\altaffilmark{A12}, Shota~Miyazaki\altaffilmark{A17},
Masayuki~Nagakane\altaffilmark{A17}, Cl\'ement~Ranc\altaffilmark{A13}, Nicholas~J.~Rattenbury\altaffilmark{A15},
Haruno~Suematsu\altaffilmark{A17}, Denis~J.~Sullivan\altaffilmark{A20}, Takahiro~Sumi\altaffilmark{A17},
Daisuke~Suzuki\altaffilmark{A21}, Paul~J.~Tristram\altaffilmark{A22}, Atsunori~Yonehara\altaffilmark{A23}\\
(The MOA Collaboration),\\
}

\email{cheongho@astroph.chungbuk.ac.kr}

\altaffiltext{A01}{Department of Physics, Chungbuk National University, Cheongju 28644, Republic of Korea} 
\altaffiltext{A02}{Institute of Natural and Mathematical Sciences, Massey University, Auckland 0745, New Zealand}
\altaffiltext{A03}{Korea Astronomy and Space Science Institute, Daejon 34055, Republic of Korea} 
\altaffiltext{A04}{Department of Astronomy, Ohio State University, 140 W. 18th Ave., Columbus, OH 43210, USA} 
\altaffiltext{A05}{Max Planck Institute for Astronomy, K\"onigstuhl 17, D-69117 Heidelberg, Germany} 
\altaffiltext{A06}{University of Canterbury, Department of Physics and Astronomy, Private Bag 4800, Christchurch 8020, New Zealand} 
\altaffiltext{A07}{Korea University of Science and Technology, 217 Gajeong-ro, Yuseong-gu, Daejeon, 34113, Republic of Korea} 
\altaffiltext{A08}{Harvard-Smithsonian Center for Astrophysics, 60 Garden St., Cambridge, MA 02138, USA} 
\altaffiltext{A09}{IPAC, Mail Code 100-22, Caltech, 1200 E.\ California Blvd., Pasadena, CA 91125, USA}
\altaffiltext{A10}{School of Space Research, Kyung Hee University, Yongin, Kyeonggi 17104, Korea} 
\altaffiltext{A11}{Department of Astronomy \& Space Science, Chungbuk National University, Cheongju 28644, Republic of Korea}
\altaffiltext{A12}{Institute for Space-Earth Environmental Research, Nagoya University, Nagoya 464-8601, Japan}
\altaffiltext{A13}{Code 667, NASA Goddard Space Flight Center, Greenbelt, MD 20771, USA}
\altaffiltext{A14}{Department of Astronomy, University of Maryland, College Park, MD 20742, USA}
\altaffiltext{A15}{Department of Physics, University of Auckland, Private Bag 92019, Auckland, New Zealand}
\altaffiltext{A16}{Okayama Astrophysical Observatory, National Astronomical Observatory of Japan, 3037-5 Honjo, Kamogata, Asakuchi, Okayama 719-0232, Japan}
\altaffiltext{A17}{Department of Earth and Space Science, Graduate School of Science, Osaka University, Toyonaka, Osaka 560-0043, Japan}
\altaffiltext{A18}{Department of Astronomy, Graduate School of Science, The University of Tokyo, 7-3-1 Hongo, Bunkyo-ku, Tokyo 113-0033, Japan}
\altaffiltext{A19}{National Astronomical Observatory of Japan, 2-21-1 Osawa, Mitaka, Tokyo 181-8588, Japan}
\altaffiltext{A20}{School of Chemical and Physical Sciences, Victoria University, Wellington, New Zealand}
\altaffiltext{A21}{Institute of Space and Astronautical Science, Japan Aerospace Exploration Agency, 3-1-1 Yoshinodai, Chuo, Sagamihara, Kanagawa, 252-5210, Japan}
\altaffiltext{A22}{University of Canterbury Mt.\ John Observatory, P.O. Box 56, Lake Tekapo 8770, New Zealand}
\altaffiltext{A23}{Department of Physics, Faculty of Science, Kyoto Sangyo University, 603-8555 Kyoto, Japan}
\altaffiltext{101}{KMTNet Collaboration.}
\altaffiltext{102}{MOA Collaboration.}

\begin{abstract}
We present the analysis of the planetary microlensing event MOA-2016-BLG-319. 
The event light curve is characterized by a brief ($\sim 3$ days) anomaly 
near the peak produced by minor-image perturbations. 
From modeling, we find two distinct solutions that describe the observed 
light curve almost equally well.  From the investigation of the lens-system configurations, 
we find that the confusion in the lensing solution is caused by the degeneracy between the 
two solutions resulting from the source passages on different sides of the planetary caustic.  
These degeneracies can be severe for major-image perturbations but it is known that they are 
considerably less severe for minor-image perturbations.  From the comparison of the lens-system 
configuration with those of two previously discovered planetary events, for which similar 
degeneracies were reported, 
we find that the degeneracies are caused by the special source trajectories that passed the 
star-planet axes at approximately right angles.  By conducting a Bayesian analysis, 
it is estimated that the lens is a planetary system in which a giant planet with a mass 
$M_{\rm p}=0.62^{+1.16}_{-0.33}~M_{\rm J}$ ($0.65^{+1.21}_{-0.35}~M_{\rm J}$) is orbiting a 
low-mass M-dwarf host with a mass $M_{\rm h}=0.15^{+0.28}_{-0.08}~M_\odot$. 
Here the planet masses in and out of the parentheses represent the masses for the 
individual degenerate solutions.
The projected host-planet separations are 
$a_\perp\sim 0.95$ au and $\sim 1.05$ au for the two solutions.  The 
identified degeneracy indicates the need to check similar degeneracies in future analyses 
of planetary lensing events with minor-image perturbations.
\end{abstract}


\section{Introduction}\label{sec:one}

Microlensing signals of planets are often described by the phrase ``a brief anomaly''
to the lensing light curve produced by the host of the planet. However, this phrase is
oversimplified because the pattern of planet-induced anomalies greatly varies depending
on the configurations of lens systems. Furthermore, planetary signals in some cases of
lens configurations can be confused with anomalies produced by other 
reasons and this induces a degeneracy problem in which multiple interpretations exist for an observed
anomaly pattern. Therefore, identifying the origins of degeneracies in various lens
configurations is important to find correct interpretations of microlensing planets by
enabling one to check similar degeneracies in future analyses.

The degeneracies in the interpretation of planetary signals are broadly classified into two
categories: ``intrinsic'' and ``accidental''. The intrinsic degeneracies are caused by the
symmetry of the lens equation, which describes the mapping from the source position
on the lens plane into the image position on the source plane. The most well known
of these is the ``close/wide'' degeneracy in which a pair of planetary models with projected
separations from the host (normalized to the angular Einstein radius $\thetae$) $s$ and
$s^{-1}$ result in very similar anomaly patterns. This degeneracy was first found by
\citet{Griest1998} for a specific case of a planetary lens system and later
extended to general binary lenses by \citet{Dominik1999} and further investigated by 
\citet{An2005}.

Accidental degeneracies, on the other hand, occur due to the fortuitous alignment of
lensing magnification patterns arising from unrelated lens configurations. 
It was pointed out by \citet{Gaudi1998}
that a subset of binary-source events, for which the flux ratio between the
binary source stars is small and the lens approaches close to the faint source
companion, 
can produce short-term anomalies, which are similar to those of planet-induced anomalies. 
These planet/binary-source degeneracies were actually found for MOA-2012-BLG-486
\citep{Hwang2013} and OGLE-2015-BLG-1459 \citep{Hwang2018c} for which the degeneracies 
were difficult to be resolved just based on the lensing light curves and could be
resolved with additional data acquired from multi-band observations. \citet{Han2008} 
pointed out another type of accidental
degeneracies in which planetary signals can be imitated by those produced by binaries
composed of roughly equal masses. Such degeneracies were demonstrated for
OGLE-2011-BLG-0526, OGLE-2011-BLG-0950/MOA-2011-BLG-336 \citep{Choi2012} and
OGLE-2015-BLG-1212 \citep{Bozza2016}. In addition, incomplete coverage 
of the planet-induced anomalies 
can cause degeneracies in interpreting anomalies 
as demonstrated in the case of
OGLE-2012-BLG-0455/MOA-2012-BLG-206 \citep{Park2014, Hwang2018a}.

\citet{Gaudi1997} (hereafter GG1997) predicted another type of accidental degeneracy, in 
which two planetary lens configurations had similar $(s, q, \alpha)$ but with the source 
passing on different sides of the caustic. Here $q$ is the planet/host mass ratio and $\alpha$
represents the angle 
between the trajectory of the source and the axis connecting the planet and its host (source trajectory angle). 
\citet{Gould1992b} argued that (under the assumption that the source passed directly over the
caustic) one could read off the values of $(s, \alpha)$ from the three \citet{Paczynski1986}
parameters of the point-lens fit $(t_0, u_0, t_{\rm E})$ and the time of the planetary
perturbation ($t_{\rm anom}$) using the relations
\begin{equation}
s-{1\over s}=\left( u_0^2+\tau_{\rm anom}^2\right)^{1/2};\qquad
\tau_{\rm anom}={t_{\rm anom}-t_0\over t_{\rm E}},
\label{eq1}
\end{equation}
and 
\begin{equation}
\tan \alpha = {u_0\over \tau_{\rm anom}}.
\label{eq2}
\end{equation}
If the caustic is relatively small, this approach is approximately accurate, even
if the source only passes near the caustic. However, GG1997 recognized that
this would lead to two slightly different solutions depending on whether the source
passes on one side of the caustic or the other. They pointed out that the degeneracy
would be severe for perturbations produced by planets with projected planet-host
separations greater than the angular Einstein radius ($s>1$, `wide' planet): ``major image
perturbations''. For ``minor-image perturbations'', which are produced by `close' planets
with $s<1$, on the other hand, it was thought that the degeneracy would be
considerable less severe. 
This is mainly because of the qualitative difference in the
caustic structures between the wide and close planetary systems, in which a wide planet
induces a single set of planetary caustics and a close planet induces two sets. 
In Appendix, we review basic facts about the types of planetary anomalies caused by major 
and minor-image perturbations for readers who are not familiar with microlensing jargon.
Experts will skip this appendix.
For the major-image caustic, the magnification pattern
on the near and far sides of the caustic are similar, and thus the anomalies produced by
the source passing both sides of the caustic are similar to each other. 
For minor-image
perturbations, on the other hand, the source trajectory passing the inner cusp will, in
general, approach close to one of the two planetary caustics, while the source trajectory
passing the outer cusp will approach the other caustic. As a result, the degeneracy would
be {\it generally} resolvable from the presence (absence) and/or timing of the anomalies
produced by the individual caustics.

Degeneracies involved with planetary caustics have been demonstrated for actual lensing
events and new types of degeneracies are additionally found with the increasing number of 
microlensing planets.  An example of the GG1997 degeneracy was recently found for the 
planetary event OGLE-2017-BLG-0173 \citep{Hwang2018d}, for which there existed 3 degenerate 
solutions and among them 2 solutions were caused by the degeneracy predicted by GG1997.  We 
note that the other solution results from a new discrete degeneracy between the solution in 
which the caustic is fully enveloped (``Cannae'' solution) and the solution in which only one 
side of the caustic is enveloped (``von Schlieffen'' solution): ``Hollywood'' degeneracy. 
The two solutions resulting from the Hollywood degeneracy have different mass ratios 
because the source passes through the caustic in different places. 
\citet{Skowron2018} found a more specific case of GG1997 degeneracies from the
analysis of the planetary lensing event OGLE-2017-BLG-0373. This so-called ``caustic-chiral''
degeneracy arises when the source passes over the caustic (contrary to the GG1997 degeneracy), but there 
are gaps in data. In this case, the solutions have very similar
$(s,\alpha)$ but substantially different $q$. Another example of the caustic-chiral degeneracy
was found by \citet{Hwang2018a} for KMT-2016-BLG-0212.
In addition, there is a case of the source passing through a major-image caustic and being 
degenerate with passing through a minor-image caustic.
This degeneracy was identified for KMT-2016-BLG-1107 by \citet{Hwang2018b}.

In this work, we analyze the planetary event MOA-2016-BLG-319.
The light curve is characterized by a short-term anomaly near the peak 
produced by the minor-image perturbation.
Contrary to the expectation that interpreting minor-image 
perturbations would not suffer from degeneracies, we find two discrete solutions that describe 
the observed light curve almost equally well.  Similar degeneracies in minor-image perturbations 
were reported for two planetary events OGLE-2016-BLG-1067 \citep{Calchi2018b} and OGLE-2012-BLG-0950 
\citep{Koshimoto2017}.  By analyzing the similarity between the anomalies of the events, we 
investigate the origin of the degeneracy.

\begin{figure}
\includegraphics[width=\columnwidth]{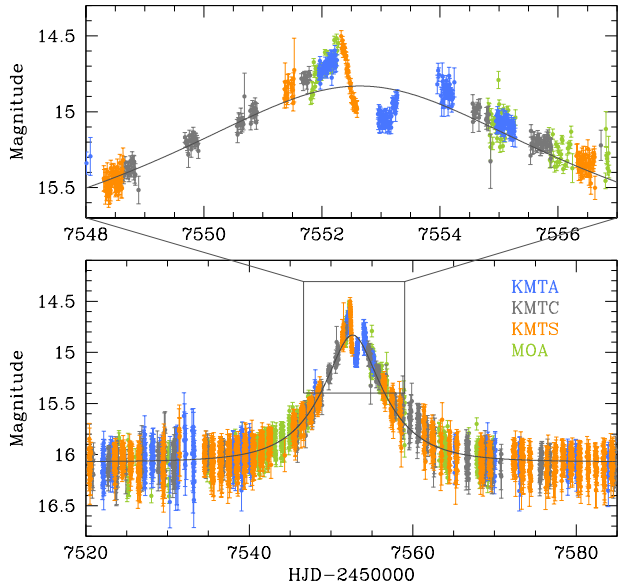}
\caption{
Light curve of the lensing event MOA-2016-BLG-319. 
The zoom of the anomaly near the peak is shown in the upper panel. 
Superposed on the data points represents the point-source point-lens model. 
\vskip0.5cm
}
\label{fig:one}
\end{figure}

\begin{figure*}
\epsscale{0.95}
\plotone{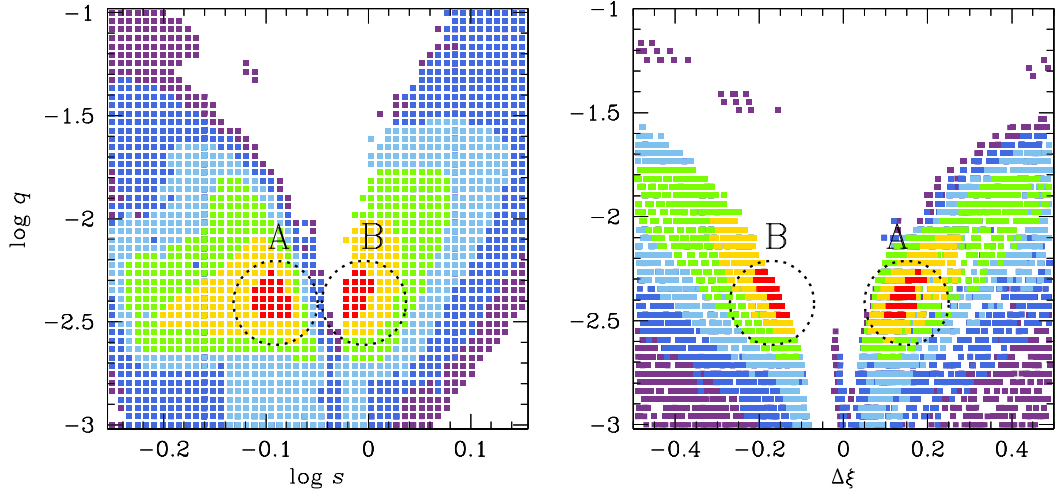}
\caption{
$\Delta\chi^2$ distributions of points in the MCMC chain on the $\log s$--$\log q$ 
(left panel) and $\Delta\xi$--$\log q$ (right panel) planes. 
The parameter $q$ denotes the planet/host mass ratio,
$s$ is the projected planet-host separation, and 
$\Delta\xi$ represents the separation between the source and the planetary caustic 
at the time of the anomaly.
Points marked in different colors represents those in the MCMC chain with 
$< 10\sigma$ (red), $< 20\sigma$ (yellow), $< 30\sigma$ 
(green), $< 40\sigma$ (cyan), $< 50\sigma$ (blue), and $< 60\sigma$ (purple).
\vskip0.3cm
}
\label{fig:two}
\end{figure*}

\section{Observations}\label{sec:two}

We identify the case of degeneracies in minor-image perturbations from the analysis of the lensing 
event MOA-2016-BLG-319.  
Figure~\ref{fig:one} shows the event light curve.
The source star of the event is located at the Galactic bulge field with equatorial coordinates 
$({\rm RA}, {\rm DEC})_{\rm J2000} = (17:54:58.13,-29:45:01.67)$. 
The corresponding galactic coordinates are $(l, b) = (0.35^\circ, -2.17^\circ)$.
The magnification of the source star induced by lensing was 
first detected and announced to microlensing community on 2016 June 13, 
${\rm HJD}^\prime={\rm HJD}-2450000\sim 7552$, by the Microlensing Observations in Astrophysics (MOA)
survey \citep{Bond2001, Sumi2003}.
The MOA survey used the 1.8~m telescope located at Mt.~John Observatory, New Zealand.
The day of the event alert approximately corresponds to not 
only the time of the light curve peak but also the start of a short-term anomaly, which lasted 
for $\sim 3$ days during $7551.5 \lesssim {\rm HJD}^\prime \lesssim 7554.5$.  However, it was difficult 
to notice the anomaly because the MOA survey could not cover the event for 4 consecutive nights before 
the anomaly and the photometry of data during the anomaly was not good enough to delineate the anomaly 
pattern. As a result, little attention was paid to the event during the progress of the event.

The scientific importance of the event was noticed with the additional data acquired by Korea Microlensing 
Telescope Network (KMTNet) survey \citep{Kim2016} conducted using three 1.6 m telescopes.
The individual KMTNet telescopes are positioned at the Cerro Tololo Interamerican Observatory, Chile (KMTC), 
the South African Astronomical Observatory, South Africa (KMTS), and the Siding Spring Observatory, 
Australia (KMTA).
The alert system of the KMTNet survey started from the 2018 season \citep{Kim2018b} and the progress of 
the event was not known in real time at the time of the event during the 2016 season.  From the analysis 
of lensing events identified by applying the Event Finder algorithm \citep{Kim2018a, Kim2018b} to the 
2016 season data, it was found that KMTNet data densely covered the light curve peak which clearly showed a 
short-term anomaly.  See the zoom of the light curve around the anomaly region presented in the upper 
panel of Figure~\ref{fig:one}.  The event was identified by KMTNet as KMT-2016-BLG-1816.

\begin{deluxetable}{lcc}
\tablecaption{Data sets used in the analysis\label{table:one}}
\tablewidth{0pt}
\tablehead{
\multicolumn{1}{c}{\ \ \ \ \ \ \ Data set\ \ \ \ \ \ \ }       &
\multicolumn{1}{c}{Range}          &
\multicolumn{1}{c}{\ \ \ \ \ \ \ $N_{\rm data}$\ \ \ \ \ \ \ }  
}
\startdata                                              
MOA            & $7502 \lesssim {\rm HJD}^\prime \lesssim 7569$     &   757  \\
KMTA (BLG02)   & $7440 \lesssim {\rm HJD}^\prime \lesssim 7676$     &   798  \\   
KMTA (BLG42)   & $7443 \lesssim {\rm HJD}^\prime \lesssim 7676$     &   701  \\   
KMTC (BLG02)   & $7439 \lesssim {\rm HJD}^\prime \lesssim 7675$     &  1143  \\   
KMTC (BLG42)   & $7439 \lesssim {\rm HJD}^\prime \lesssim 7675$     &  1006  \\   
KMTS (BLG02)   & $7441 \lesssim {\rm HJD}^\prime \lesssim 7681$     &  1394  \\   
KMTS (BLG42)   & $7441 \lesssim {\rm HJD}^\prime \lesssim 7681$     &  1472    
\enddata                            
\tablecomments{ 
${\rm HJD}^\prime={\rm HJD}-2450000$.
}
\end{deluxetable}

In Table~\ref{table:one}, we present the data sets used in the analysis.  MOA observations 
of the event was conducted in a customized $R$ band with a cadence of 1 hr.  KMTNet observations 
were conducted mostly in $I$ band with occasional observations in $V$ band 
to measure the source color. 
The event was in the KMTNet BLG02 and BLG42 fields which were monitored with a 
cadence of 0.5 hr by the individual telescopes. For the period from April 23 (${\rm HJD}'\sim 2457501$) 
to June 16 (${\rm HJD}'\sim 2457555$), the cadence of KMTS and KMTA was increased in order to 
support {\it Kepler} K2 C9 campaign \citep{Gould2013}. While the event does not lie in the K2 field, 
the anomaly coverage serendipitously benefited from this cadence increase.  
The columns `range' and 
`$N_{\rm data}$' indicate the time range of the data sets used for analysis and the number of 
data points constituting the individual data sets, respectively.
We set the range of the MOA data in the region around event 
because the baseline data exhibit considerable fluctuation.

Photometry of data are processed using the codes of the individual groups: 
\citet{Bond2001} for the MOA survey and \citet{Albrow2009} for the KMTNet survey.
Both codes utilize the difference imaging method developed by \citep{Alard1998}.
Errorbars are normlized using the recipe explained in \citet{Yee2012}.

\section{Light curve Analysis}\label{sec:three}

The light curve shows a pronounced dip.  Such a dip feature in lensing light curves
can only be produced by minor image
perturbations.  There are two possibilities in the lens-system configuration.
First, minor image gives rise to two triangular planetary caustics with a magnification
dip between them.
Second, there
is a six-sided resonant caustic whose ``back end'' 
consists of two caustic wings separated by a dip.
See Figure 4 of \citet{Gaudi2012} for the variation of planetary microlens caustics.

Although some lens configurations are excluded in advance based on the previously well-studied 
origins of degeneracies, interpreting the anomaly may be subject to unknown types of degeneracies. 
We, therefore, conduct a thorough 
grid search for the planetary lensing parameters $s$ and $q$. 
Besides these planetary parameters, one needs additional lensing parameters to model the 
observed light curve. 
These parameters describe the source star's approach to the lens including 
the time of the closest approach, $t_0$, the lens-source separation at that time, 
$u_0$ (impact parameter), the event time scale, $t_{\rm E}$ (Einstein time scale), and the 
source trajectory angle. 
The anomaly might be produced by the crossings of the source over caustics. 
There is no obvious 
signature of caustic crossings, which usually produce sharp spike feature.  However, 
caustic-crossing features can be smooth if the source is substantially bigger than the caustic. 
Even if a source is smaller than an overall caustic, it could be big compared to the 
caustic figure that it is passing over, e.g., 
OGLE 2016-BLG-1195 \citep{Bond2017} and OGLE-2016-BLG-1195 \citep{Shvartzval2017}.
These finite-source effects were theoretically predicted by 
\citet{Bennett1996} and observationally confirmed by \citet{Beaulieu2006} for the 
planetary lensing event OGLE-2005-BLG-390. 
To account for possible finite-source effects, we 
include an additional parameter of the normalized source radius $\rho$, 
which denotes the ratio of the angular source radius $\theta_*$ to 
the angular Einstein radius $\thetae$, i.e., $\rho=\theta_*/\thetae$.
For a given set of the planetary parameters $s$ and $q$, 
we search for the other parameters using the Markow Chain Monte Carlo (MCMC) method.
We set the ranges of the grid parameters, i.e., $s$ and $q$, wide 
enough to check the possibility that the anomaly is produced by 
binaries that have similar mass components.

In the left panel of Figure~\ref{fig:two}, we present $\Delta\chi^2$ distribution of 
points in the MCMC chain on the $\log s$--$\log q$ plane 
acquired from the preliminary grid search.  
From the distribution, one first finds that 
the lens responsible for the anomaly is composed of two masses with 
a very low mass ratio of $q\sim 4\times 10^{-3}$, 
suggesting that the lower-mass 
component of the lens is a {\it planet}. One also finds that there exist two distinct solutions 
centered at $(\log s, \log q)\sim (-0.09, -2.4)$ (marked by ``A'') and $\sim (-0.02, -2.4)$ 
(``B''). From further refinement of the individual local solutions by letting all parameters 
vary, it is found that the $\chi^2$ difference between the two solutions is merely 
$\Delta\chi^2\sim 1.9$. This indicates that both solutions describe the observed anomaly 
almost equally well, although the solution ``A'' is slightly preferred over the solution 
``B''.

\begin{deluxetable}{lcc}
\tablecaption{Best-fit lensing parameters \label{table:two}}
\tablewidth{0pt}
\tablehead{
\multicolumn{1}{c}{\ \ \ \ \ \ \ Parameter\ \ \ \ \ \ \ } &
\multicolumn{1}{c}{Inner solution (``A'')}   &
\multicolumn{1}{c}{Outer solution (``B'')}  
}
\startdata                                              
$\chi^2$                   &  7308.1                  &  7310.0                  \\
$t_0$   (HJD$^\prime$)     &  7552.737 $\pm$ 0.013    &  7552.742 $\pm$ 0.012    \\   
$u_0$                      &     0.267 $\pm$ 0.012    &     0.260 $\pm$ 0.012    \\   
$t_{\rm E}$ (days)         &     8.60  $\pm$ 0.26     &     8.69  $\pm$ 0.26     \\   
$s$                        &     0.817 $\pm$ 0.004    &     0.945 $\pm$ 0.008    \\   
$q$ ($10^{-3}$)            &     3.93  $\pm$ 0.11     &     4.10  $\pm$ 0.12     \\   
$\alpha$ (rad)             &     4.646 $\pm$ 0.006    &     4.645 $\pm$ 0.006    \\  
$F_{\rm s,KMTC}$           &     6.25  $\pm$ 0.08     &     6.23  $\pm$ 0.08     \\   
$F_{\rm b,KMTC}$           &    -0.10  $\pm$ 0.11     &    -0.10  $\pm$ 0.11      
\enddata                            
\tablecomments{ 
${\rm HJD}^\prime={\rm HJD}-2450000$.
}
\end{deluxetable}

To be noted is that the degeneracy between the two solutions is different from the previously 
known `close/wide' degeneracy.  The two solutions resulting from the close/wide degeneracy
have planet-host separations $s$ and 
$s^{-1}$ and thus one solution has a separation smaller than $\thetae$, i.e., $s<1$,  and 
the other solution has a separation greater than $\thetae$, $s>1$. In the case of MOA-2016-BLG-319, 
both degenerate solutions have separations $s<1$ ($s\sim 0.82$ for the solution ``A'' and 
$s\sim 0.95$ for the solution ``B'') indicating that the origin of the degeneracy 
does not stem from the symmetry of the lens equation.\footnote{We note that the red zone 
in the $\Delta\chi^2$ distribution
(presented in the left panel of Fig.~\ref{fig:two}) covers $s=1$ and even
slightly greater. If $s\sim 1$, then the lens system forms a
resonant caustic, in which the central and planetary caustic merge together.
In this case, the back-end of the resonant caustic still
induce a dip in the light curve, even if the binary separation
$s$ is greater than unity.
}

In Table~\ref{table:two}, we present the best-fit lensing parameters for the two degenerate 
solutions.
We also present the $\chi^2$ values of the fit for the solutions.
The uncertainties of the parameters correspond to 
the scatter of points in the MCMC chain.
We note that the 
lensing parameters of the two degenerate solutions are very similar to each other except 
for the binary separation $s$.  To be also noted is that the time scale of the event, 
$t_{\rm E}\sim 8.6$ days, is short.  As a result, higher-order effects  
induced by the orbital motion of the Earth, microlens-parallax effect \citep{Gould1992a}, 
or that of the lens, lens-orbital effect \citep{Dominik1998}, 
is not important in describing the observed light curve.
Also presented in the table are the flux values of the source, $F_{\rm s,KMTC}$,
and blend, $F_{\rm b,KMTC}$, as measured from the pyDIA photometry of the KMTC data set. 
We note that the blend flux has a slightly negative value but it is
consistent to be zero within  the measurement error. 
These measured values of $F_{\rm s,KMTC}$
and $F_{\rm b,KMTC}$ indicate that the flux from the source dominates the blended flux.

\begin{figure}
\includegraphics[width=\columnwidth]{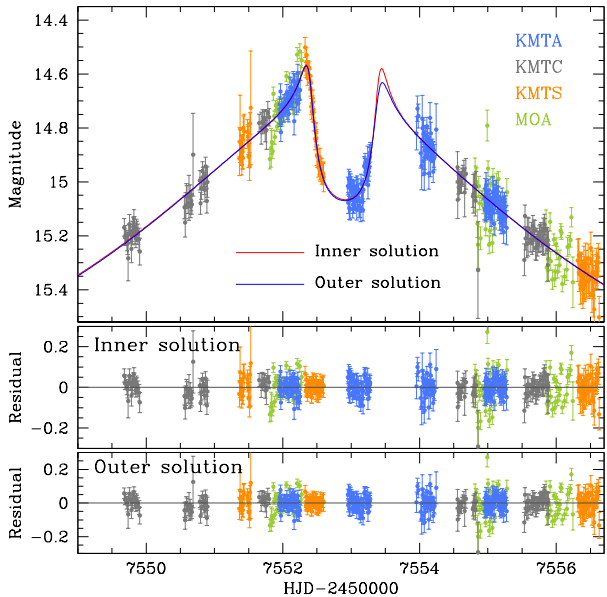}
\caption{
Two model light curve resulting from the two degenerate solutions: ``inner solution'' and 
``outer solution''. The lower panels show the residuals from the individual solutions.  The 
lens-system configurations corresponding to the individual solutions 
are presented in Fig.~\ref{fig:four}.
\vskip0.3cm
}
\label{fig:three}
\end{figure}

In Figure~\ref{fig:three}, we present model light curves for both solutions plotted over 
the observed data points. Except for the very short period around ${\rm HJD}^\prime \sim 7553.4$, 
the two model light curves are so similar to each other that it is difficult to distinguish them 
within the line width, indicating that the degeneracy between the two solutions is very severe. 
This can be also seen in the lower two panels in which the residuals from the individual solutions 
are presented.

\begin{figure}
\includegraphics[width=\columnwidth]{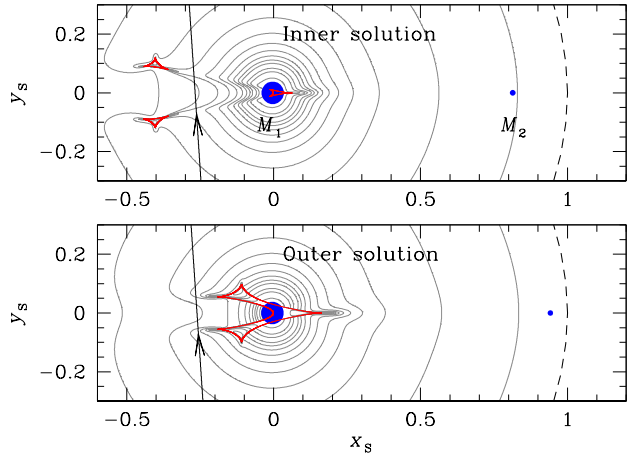}
\caption{
Lens-system configurations for the two degenerate solutions seen on the {\it source plane}.
For each panel, the curve with an arrows represents the source trajectory and the red closed 
curves represent caustics. The two dots marked by $M_1$ and $M_2$ denote the lens components, 
where $M_1 > M_2$.  The dashed circle centered at $M_1$ represents the Einstein ring.  Contours 
of lensing magnification are drawn to show the region of anomaly around caustics. 
\vskip0.3cm
}
\label{fig:four}
\end{figure}

Figure~\ref{fig:four} shows the configurations of the lens system for the individual solutions.
In each panel, the line with an arrow represents the source trajectory,
the small dots marked by $M_1$ and $M_2$ indicate the 
positions of the lens components, and 
the closed figures composed of concave curves represent the caustic. 
The upper and lower panels are the 
configurations for the solutions ``A'' and ``B'', respectively.  We draw contours of magnification 
to show the region of anomaly around the caustics. 
For the solution ``A'', the caustic is composed of 3 sets in which 
the small caustic located close to the host is the central caustic and 
the two caustics located away from the host are the planetary caustics. 
According to this solution, 
the anomaly was produced by the passage of the source 
through the region between the central and planetary caustics.
The negative 
deviation of the anomaly was produced during the time when the source passed the negative perturbation 
region between the central and planetary caustics.  Because the source trajectory passed the inner 
region of the planetary caustic with respect to the planet host, we refer to this solution as  
``inner solution''.  For solution ``B'', on the other hand, 
the lens system produces a single resonant caustic. 
This results from 
the merging of the central and planetary caustics because of the proximity of the planet-host 
separation to unity, $s\sim 0.95$. According to this solution, the source trajectory passed the 
back-end of the caustic and 
the negative deviation occurred during the time when the 
source passed the region extending from the caustic end.  
Because the source trajectory passed the outer region of the caustic (with respect to the planet host), 
we refer to this solution as ``outer solution''.

\begin{figure}
\includegraphics[width=\columnwidth]{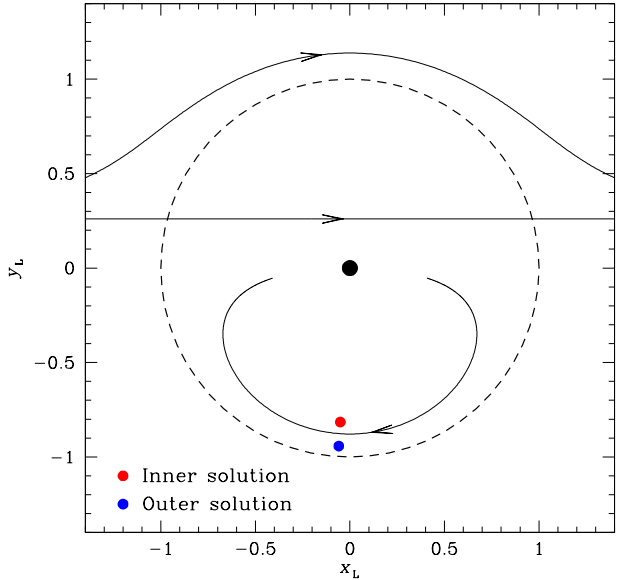}
\caption{
Lensing system configurations for the two degenerate solutions seen on the {\it lens plane}.  
The coordinates are centered at the position of the planet host (black dot) and
the dashed circle represents the Einstein ring.
The solid line with an arrow represents the path of the source.
The two solid curves with arrows represent the paths of the images produced by 
the planet host.  The red and blue dots represent the planet positions for the inner and 
outer solutions, respectively.  
\vskip0.3cm
}
\label{fig:five}
\end{figure}

For better understanding the origin of the degeneracy between the two solutions, in 
Figure~\ref{fig:five}, we present the lens-system configurations seen on the {\it lens plane}.  
In the plot, the planet host is located at the origin and the line with an arrow represents 
the path of the source.  The two solid curves with arrows represent the paths of the images 
produced by the host. The red and blue dots represent the planet positions for the inner and 
outer solutions, respectively.  From the configurations, it is found  that, for both solutions, 
the planet is located close to the minor image produced by the host at the time of the perturbation.  
The difference in the configurations between the two solutions is that the planet is located inside 
the minor image (with respect to the host) for the ``inner solution'', while it is located outside 
of the image for the ``outer solution''.  This indicates that the similarity between the two 
solutions is caused by the degeneracy in minor-image perturbations.  We note that the lens 
system configuration is similar to that presented in Figure~1 of GG1997 except that 
the positions of planets are different.

The fact that the two solutions are originated from the GG1997 degeneracy 
can also be seen in the right panel of Figure~\ref{fig:two}, in which 
we plot $\Delta\chi^2$ distribution of MCMC points on the $\Delta\xi$--$\log q$ plane.
Here $\Delta\xi$ represents the separation between the source and the planetary caustic 
at the time of the anomaly. 
The separation $\Delta\xi$ is determined from the lensing parameters by
\begin{equation}
\Delta\xi = {u_0 \over \sin \alpha} - \left( s-{1\over s}\right), 
\label{eq3}
\end{equation}
where the former term on the right side, i.e., $u_0/\sin\alpha$, represents the 
separation between the source and planet host at the time of the anomaly 
and the latter term, i.e., $s-1/s$, denotes the separation between the 
caustic center and the host. 
We note that similar plots are presented in Figure 4 of \citet{Hwang2018d} 
and Figure 5 of \citet{Skowron2018}. 
From the plot, it is found that 
the degenerate solutions 
have similar separations $|\Delta\xi|\sim 0.15$ 
but with opposite signs,
indicating that the source stars of the individual solutions 
approach the opposite sides of the caustic with similar separations 
from the caustic.

Despite that degeneracies are thought to be considerably less severe for minor-image
perturbations compared to major-image perturbations, similar degeneracies in minor-image 
perturbations were reported by \citet{Calchi2018b} and \citet{Koshimoto2017} for the 
planetary events OGLE-2016-BLG-1067 and OGLE-2012-BLG-0950, respectively.  We, therefore, 
compare the lens-system configurations of MOA-2016-BLG-319 with the two other events in order 
to find the cause of the degeneracy.  The lens-system configurations of OGLE-2016-BLG-1067 
are presented in Figures 3 and 4 of \citet{Calchi2018b}.  We note that they presented 8 
degenerate configurations, among which a four-fold degeneracy is caused by the space-based 
parallax measurement \citep{Refsdal1966, Gould1994} and the other two-fold degeneracy is 
relevant to the minor-image perturbation.  The 
configurations
of OGLE-2012-BLG-0950 are 
presented in Figure~2 of \citet{Koshimoto2017}.

From comparing the lens-system configurations of the events, we find one major difference 
and one major similarity.  The difference is that the planetary caustics of the outer 
solution is separated from the central caustic for OGLE-2016-BLG-1067, while it is merged 
with the central caustic for MOA-2016-BLG-319 and OGLE-2012-BLG-0950 (resonant topology).  
However, in the sense that the source trajectory of MOA-2016-BLG-319 and OGLE-2012-BLG-0950 
passed the planetary wing of the resonant caustic, the degeneracies of the events are 
considered to be of the same type.  The similarity is that the source stars of the events 
passed the planet-host axes at about right angles.  The source trajectory angles are 
$\alpha\sim 267^\circ$, $\sim 278^\circ$, $\sim 292^\circ$ for MOA-2016-BLG-319, 
OGLE-2016-BLG-1067, and OGLE-2012-BLG-0950, respectively.  For general cases in which the 
entrance angle of the source is substantially different from a right angle, the source 
trajectory passing the inner region of the minor-image caustic produces a light curves that 
is distinct from a trajectory passing the outer region.  On one side of the set of caustics, 
they will pass closer to one of the two caustics and farther from the other caustic. On the 
other side, the order will be inverted (for fixed angle).  This will result in different 
anomaly patterns, and thus one can easily distinguish the two cases if the anomaly is densely 
covered.  In the case of a right angle source entrance, on the other hand, the source approaches 
the individual caustics with approximately same distances.  Then, the anomaly patterns resulting 
from the source trajectories passing the inner and outer regions of the planetary caustic can 
appear to be similar.  We, therefore, conclude that the degeneracies in the minor-image 
perturbations of both events occur because the source stars crossed the star-planet axes at 
approximately right angles.

\begin{figure}
\includegraphics[width=\columnwidth]{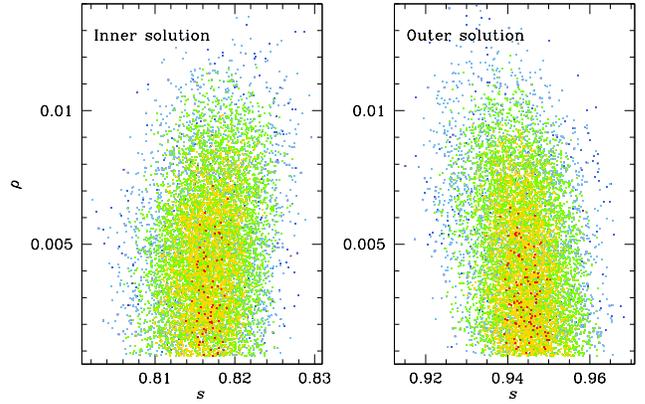}
\caption{
Distributions of $\Delta\chi^2$ of points in the MCMC chain on the 
$s$--$\rho$ parameter plane.
Color coding represents points within $1\sigma$ (red), $2\sigma$ (yellow), $3\sigma$ 
(green), $4\sigma$ (cyan), and $5\sigma$ (blue).
The left and right panels are the distributions for the inner and outer 
solutions, respectively.
\vskip0.3cm
}
\label{fig:six}
\end{figure}

\section{Source star}\label{sec:four}

We characterize the source star based on the source flux measured in $I$ and $V$ passbands.  
Besides simply knowing the type of the source star, 
characterizing the source star is important 
because it may provide information about the angular Einstein radius 
in combination with 
the normalized source radius $\rho$ via the relation $\thetae=\theta_*/\rho$.
For MOA-2016-BLG-319, however, the normalized source radius $\rho$ cannot be measured 
because deviations in the lensing light curve 
caused by finite-source effects cannot be firmly detected.
See Figure~\ref{fig:six}, in which we plot the $\Delta\chi^2$ distributions of points in the MCMC 
chain on the $s$--$\rho$ parameter plane.  However, the upper limit on $\rho$ can be measured 
and this yields 
a lower limit on $\thetae$, 
which may provide a constraint on the physical lens parameters.
It is found that the upper limit of the normalized source radius 
is $\rho_{\rm max}\sim 0.01$ as measured at the $3\sigma$ level.

\begin{figure}
\includegraphics[width=\columnwidth]{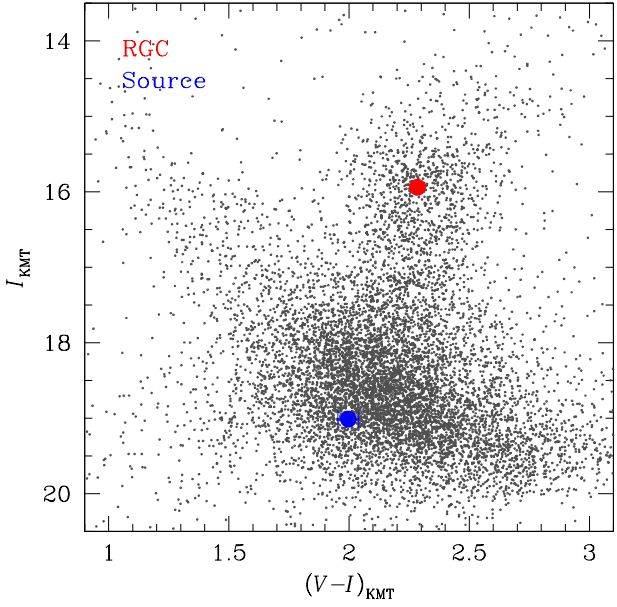}
\caption{
Position of the source star in the instrumental color-magnitude diagram constructed 
based on the KMTNet data. Also marked is the location of red giant clump (RGC) centroid.
\vskip0.3cm
}
\label{fig:seven}
\end{figure}

The de-reddened color $(V-I)_0$ and brightness $I_0$ of the source star
are estimated 
using the known values of the red giant clump (RGC) centroid,
$(V-I,I)_{{\rm RGC},0}=(1.06,14.41)$ \citep{Bensby2011, Nataf2013},
and the offsets in color and magnitude of the source from the RGC centroid. 
In Figure~\ref{fig:seven}, we mark the positions of the source and RGC centroid in the 
color-magnitude diagram of stars 
located in the same field of the source.
The positions of the source and the RGC centroid are
$(V-I,I)=(2.00\pm 0.04, 19.01\pm 0.01)$ and  
$(V-I,I)_{\rm RGC}=(2.29, 15.94)$, respectively.
From the color and brightness offsets, 
it is found that the de-reddened color and brightness of the source star are 
$(V-I,I)_0=(0.77\pm 0.04, 17.49\pm 0.01)$, 
respectively.  
The estimated de-reddened color and magnitude of the source star indicate that 
the source is likely to be a turnoff star.
We then convert the measured $V-I$ color
into $V-K$ color using the $(V-I)/(V-K)$ relation  of \citet{Bessell1988}. 
Finally, we estimate the 
source angular radius using the $(V-K)/\theta_*$ relation of \citet{Kervella2004}.  
The estimated 
angular source radius is 
\begin{equation}
\theta_*= 1.07\pm 0.09\ \mu{\rm as}.  
\label{eq4}
\end{equation}
With the measured angular source radius, the lower limit of the angular Einstein radius is 
set to be
\begin{equation}
\theta_{\rm E,min}={\theta_* \over \rho_{\rm max}}\sim 0.107~{\rm mas}.
\label{eq5}
\end{equation}

\section{Bayesian Analysis of Lens Parameters}\label{sec:five}

In order to uniquely determine the mass, $M$, and distance to the lens, $D_{\rm L}$,
it is required to measure both the microlens parallax $\pie$ and the angular Einstein 
radius $\thetae$, from which $M$ and $D_{\rm L}$ are determined by
\begin{equation}
M={\thetae \over \kappa \pie};\qquad
D_{\rm L}={{\rm au} \over \pie\thetae + \pi_{\rm S} }
\label{eq6}
\end{equation} 
\citep{Gould2000b}.
Here $\kappa=4G/(c^2{\rm au})\sim 8.14~{\rm mas}~M_\odot^{-1}$ and $\pi_{\rm S}={\rm au}/D_{\rm S}$ denotes the parallax 
of the source located at a distance $D_{\rm S}$.  For MOA-2016-BLG-319, however, neither 
$\pie$ nor $\thetae$ is measured, although the upper limit of $\thetae$ is set.  We, 
therefore, conduct a Bayesian analysis of Galactic lensing events to estimate $M$ and 
$D_{\rm L}$ based on the measured event time scale. 
The time scale provides a constraint on the lens parameters 
because it is related to the parameters by 
\begin{equation}
t_{\rm E}={\sqrt{\kappa M \pi_{\rm rel}} \over \mu};\qquad
\pi_{\rm rel}={\rm au} \left( {1\over D_{\rm L}}-{1\over D_{\rm S}}\right).
\label{eq7}
\end{equation}
Here $\mu$ represents the relative lens-source proper motion.  We also use the constraint 
of $\theta_{\rm E,min}$.

Implementing a Bayesian analysis requires models 
describing how lens objects are distributed, i.e., physical distribution, and how they move,
i.e., dynamical distribution.
One also needs a model mass function of lens objects. 
We construct the lens mass function based on the \citet{Chabrier2003} mass function for stars 
combined with the \citet{Gould2000a} mass function for stellar remnants including black holes, 
neutron stars, 
and white dwarfs.  Lens and source objects are assumed to be distributed following the 
physical matter distribution model of \citet{Han2003}, in which 
the disk has a  double-exponential form and the 
bulge has a triaxial shape. 
For the dynamical distribution, we adopt \citet{Han1995} model, in which the 
motion of disk objects follows a gaussian velocity distribution 
with a mean corresponding to the rotation speed of the disk, 
and the motion of bulge objects follows a triaxial gaussian distribution 
with the velocity components along the individual axes 
deduced from the bulge shape via the tensor virial theorem.
Based on the model distributions, we conduct a Monte Carlo simulation to generate a large 
number ($6\times 10^6$) of Galactic lensing events.  
Then, the the lens mass and distance distributions are constructed 
based on the events with time scales located 
within the range of the measured event time scale.
With these distributions, we then estimate the representative values of $M$ and 
$D_{\rm L}$ as the median values.
The lower and upper limits of the values are estimated as 
the 16\% and 84\% of the distribution.

\begin{figure}
\includegraphics[width=\columnwidth]{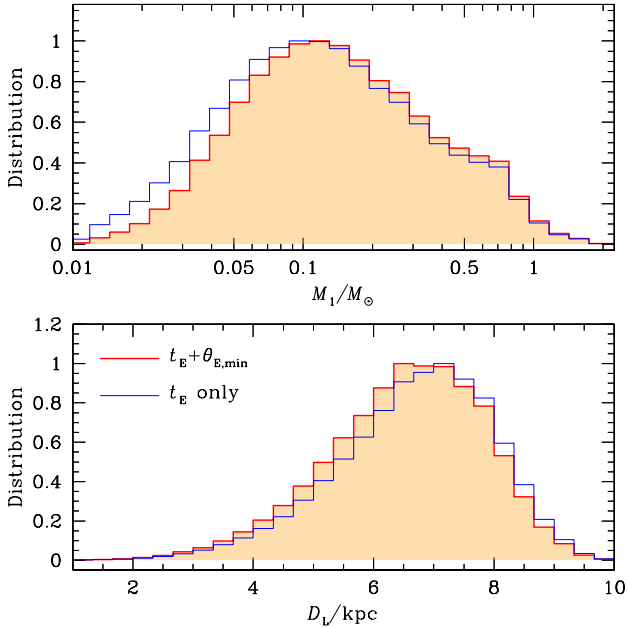}
\caption{
Probability distributions of the mass (upper panel) and distance (lower panel) of the 
planet host.  The blue curve is the distribution based on only the event time scale 
$t_{\rm E}$, while the red curve is the distribution with the additional constraint of 
the lower limit of the angular Einstein radius, $\theta_{\rm E,min}$. 
\vskip0.3cm
}
\label{fig:eight}
\end{figure}

In Figure~\ref{fig:eight}, we present the distributions of the lens mass 
(upper panel) and distance (lower panel) obtained from the Bayesian analysis.
In each panel, the blue curve is the distribution 
based on only the event time scale $t_{\rm E}$, while the red curve is the distribution 
obtained with 
the additional constraint of $\theta_{\rm E,min}$.  It is found that the constraint of 
$\theta_{\rm E,min}$ is weak and thus has little effect on the probability distribution.  
The estimated masses of the lens components are
\begin{equation}
M_1=0.15^{+0.28}_{-0.08}~M_\odot
\label{eq8}
\end{equation}
and
\begin{equation}
M_2=0.62^{+1.16}_{-0.33}~M_{\rm J}.
\label{eq9}
\end{equation}
Therefore, the lens is a planetary system in which a giant planet is orbiting 
a low-mass M dwarf.  Planetary systems belonging to low-mass hosts are difficult to be
detected by other major planet detection methods, e.g., radial velocity or transit methods, 
due to the faintness of host stars.  On the other hand, the microlensing method does not rely on 
the brightness of the host star, and thus the majority of planetary systems with low-mass stars 
were found using the microlensing method.  See Figure~10 of \citet{Han2018} and Figure~6 of 
\citet{Jung2018}, which show the distribution of planets in 
the $M_1$--$M_2$ plane.
The planetary system is estimated to be located at a distance of
\begin{equation}
D_{\rm L} = 6.8^{+1.2}_{-1.4}~{\rm kpc}.
\label{eq10} 
\end{equation}

In Table~\ref{table:three}, we list the physical lens parameters.
We note that the 
ranges of the lens mass and distances are considerable due to the Bayesian nature of 
determining the lens parameters combined with weak constraint of extra information, e.g., 
$\thetae$ or $\pie$.  Also presented in the table is the projected separation between the 
planet and host, which is estimated by $a_\perp = s D_{\rm L}\thetae$.  The projected 
separation is $a_\perp\sim 0.95$ au and $\sim 1.05$ au for the inner and outer solutions, 
respectively.  In both cases, the planet is located away from the snow line at 
$a_{\rm snow}\simeq 2.7~{\rm au}(M/M_\odot)\sim 0.4~{\rm au}$.

\begin{deluxetable}{lcc}
\tablecaption{Physical Lens Parameters \label{table:three}}
\tablewidth{0pt}
\tablehead{
\multicolumn{1}{c}{\ \ \ \ \ \ \ Parameter\ \ \ \ \ \ \ } &
\multicolumn{1}{c}{\ \ \ \ \ \ \ Inner solution\ \ \ \ \ \ \ }   &
\multicolumn{1}{c}{\ \ \ \ \ \ \ Outer solution\ \ \ \ \ \ \ }   
}
\startdata                                              
$M_1$        ($M_\odot$)   & $0.15^{+0.28}_{-0.08}$    &  $0.15^{+0.28}_{-0.08}$  \\
$M_2$        ($M_{\rm J}$) & $0.62^{+1.16}_{-0.33}$    &  $0.65^{+1.21}_{-0.35}$  \\
$D_{\rm L}$  (kpc)         & $6.8^{+1.2}_{-1.4}$       &  $6.8^{+1.2}_{-1.4}$     \\
$a_\perp$    (au)          & $0.95^{+0.17}_{-0.20}$    &  $1.09^{+0.19}_{-0.22}$   
\enddata                            
\end{deluxetable}

\section{Summary}\label{sec:six}

We presented the analysis of the planetary lensing event MOA-2016-BLG-319
for which the light curve was characterized by a short-term anomaly near the peak produced 
by the minor-image perturbation.
From modeling of the light curve, we found that there 
existed two distinct solutions that described the observed light curve almost equally well.  
The planet-host separations of both solutions were smaller than the Einstein radius,
indicating that 
the degeneracy was different from the previously known `close/wide' degeneracy.  From the 
investigation of the lens-system configurations, it was found that the two solutions resulted 
from the degeneracy in minor-images perturbations.  Because it had been believed that the 
degeneracy in determining the planet parameters would not be severe for 
minor-image perturbations, such a degeneracy was unexpected.
From the comparison 
of the lens-system configuration with those of OGLE-2016-BLG-1067 and OGLE-2012-BLG-0950, 
for which similar degeneracies
were reported, we found that the degeneracies for the events were caused by the special 
source trajectories that passed the star-planet axis at approximately right angles.
By conducting a Bayesian analysis, we estimated 
that the lens was a planetary system in which a giant planet with a mass 
$M_{\rm p}=0.62^{+1.16}_{-0.33}~M_{\rm J}$ ($0.65^{+1.21}_{-0.35}~M_{\rm J}$) was orbiting 
a low-mass M dwarf with a mass 
$M_{\rm h}=0.15^{+0.28}_{-0.08}~M_\odot$, 
where the planet masses in and out of the parentheses represent the masses for the inner and outer
solutions, respectively. 
The projected host-planet separations were 
$a_\perp\sim 0.95$ au and $\sim 1.05$ au for the individual degenerate solutions. 
The identified degeneracy indicated the need to check similar degeneracies in future analysis of 
planetary events with minor-image perturbations.

\appendix

\section{Types of Planetary Perturbations}\label{appendix}

When a source star is microlensed, 
the image of the source splits into two.
One image with a higher magnification (``major image'') 
appears outside the Einstein ring
and the other image with a lower magnification (``minor image'') 
appears inside the ring.
Planetary perturbations occur 
at the time when the planet is positioned near one of the 
two microlens images of the primary star 
and additionally perturbs the nearby image 
\citep{Mao1991, Gould1992b}.
\color{black}

Depending on which image is perturbed by the planet, planetary perturbations are classified into 
two types: ``major-image perturbation'' and ``minor-image perturbation'' (GG1997). 
The major-image perturbation indicates the case in which the major image is 
perturbed by the planet.
Because the major image is located outside the Einstein ring, major-image perturbations are caused 
by planets 
with separations $s>1$.
On the other hand, 
minor-image perturbation indicates the case in which the minor image is perturbed by the planet. 
The minor image is located inside the Einstein ring and thus minor-image perturbations 
are caused by planets 
with separations $s<1$.
When a
major image is perturbed, the image is further magnified by the planet and thus the lensing 
light curve always exhibits positive deviations from the light curve produced by the host. In 
contrast, the minor-image perturbation causes the demagnitification of the image, 
producing, in most cases, a negative deviation in the light curve.

In the view point on the source plane, planetary lensing signals are produced when a source 
approaches the caustic produced by the planet. Planets induce one or two sets of ``planetary 
caustics'' depending on 
whether the planetary separation is greater or smaller than the Einstein radius.
\footnote{We note 
that planets also induce ``central caustics'' in the region close to the host of the planet. Due to 
the location of the caustic, planetary signals produced by central caustics always occur near the 
peak of lensing events with very high magnifications \citep{Griest1998}. For the dependency of the 
location, size, and shape of the central caustic on the star-planet separation $s$ and the planet/star 
mass ratio $q$, see \citet{Chung2005}.}   For the lens system with a wide planet, there exists a single 
set of planetary caustics with four cusps. For the system with a close planet, on the other hand, 
there exist two 
sets of caustics in which one is located above the planet-host axis and the other is located below the 
axis and each of the caustics is composed of three cusps. 
For the detailed properties of planetary caustics, 
see \citet{Han2006}.

\acknowledgments
Work by CH was supported by the grant (2017R1A4A1015178) of National Research Foundation of Korea.
Work by AG was supported by US NSF grant AST-1516842.
Work by IGS and AG were supported by JPL grant 1500811.
This research has made use of the KMTNet system operated by the Korea
Astronomy and Space Science Institute (KASI) and the data were obtained at
three host sites of CTIO in Chile, SAAO in South Africa, and SSO in
Australia.
The MOA project was supported by JSPS KAKENHI Grant Number JSPS24253004,
JSPS26247023, JSPS23340064, JSPS15H00781, and JP16H06287.
DPB, AB, and CR were supported by NASA through grant NASA-80NSSC18K0274. 
The work by CR was supported by an appointment to the NASA Postdoctoral Program at the Goddard 
Space Flight Center, administered by USRA through a contract with NASA. NJR is a Royal Society 
of New Zealand Rutherford Discovery Fellow.
We acknowledge the high-speed internet service (KREONET)
provided by Korea Institute of Science and Technology Information (KISTI).

\end{document}